\documentclass[aps,prx,twocolumn,superscriptaddress,nofootinbib]{revtex4-2}
\usepackage[T1]{fontenc}
\usepackage{amssymb,graphicx,color}
\usepackage[intlimits]{amsmath}
\usepackage[english]{babel}
\usepackage[colorlinks]{hyperref}
\hypersetup{
    colorlinks=blue,
    linkcolor=blue,
    filecolor=blue,
    urlcolor=blue,
    citecolor=blue
}
\pdfoutput=1
\graphicspath{{figures/}} 
\usepackage[normalem]{ulem}
\usepackage{comment}
\usepackage{ragged2e}
\usepackage{enumerate}
\usepackage{braket}

\usepackage[dvipsnames]{xcolor}

\begin{document}

\title{\hspace{-17pt} Many-body ergodicity breaking from wavefunction snapshots}

\author{Riccardo Andreoni}
\affiliation{International School for Advanced Studies (SISSA), via Bonomea 265, 34136 Trieste, Italy}
\affiliation{The Abdus Salam International Centre for Theoretical Physics (ICTP), Strada Costiera 11, 34151 Trieste, Italy}

\author{Devendra Singh Bhakuni}
\affiliation{The Abdus Salam International Centre for Theoretical Physics (ICTP), Strada Costiera 11, 34151 Trieste, Italy}

\author{Marcello Dalmonte}
\affiliation{The Abdus Salam International Centre for Theoretical Physics (ICTP), Strada Costiera 11, 34151 Trieste, Italy}
\affiliation{Dipartimento di Fisica e Astronomia, Università di Bologna, via Irnerio 46, I-40126 Bologna, Italy}
\affiliation{INFN, Sezione di Bologna, via Irnerio 46, I-40126 Bologna, Italy}

\author{Lev Vidmar}
\affiliation{Department of Theoretical Physics, J. Stefan Institute, SI-1000 Ljubljana, Slovenia}
\affiliation{Department of Physics, Faculty of Mathematics and Physics, University of Ljubljana, SI-1000 Ljubljana, Slovenia\looseness=-1}

\author{Miroslav Hopjan}
\affiliation{Institute of Theoretical Physics, Wrocław University of Science and Technology, 50-370 Wrocław, Poland}

\begin{abstract}
Modern quantum experiments can probe many-body wavefunctions through projective measurements, providing snapshots of individual many-body configurations. A fundamental question is whether the intrinsic structure of their measurement distributions can reveal quantum dynamics beyond predefined observables. Here, we address this question at the many-body ergodicity-breaking transition using two complementary characteristics of nonequilibrium wavefunction snapshots: their binary intrinsic dimension (BID) and the topology of wavefunction networks constructed from Hamming distances. We show that the critical point is characterized by an extensive but submaximal BID, and scale-free network connectivity with highly connected hubs. These signatures distinguish the ergodic, critical, and nonergodic regimes, and provide experimentally accessible probes of ergodicity breaking from projective measurement snapshots.
\end{abstract}

\maketitle

\section{Introduction}
\label{sec:intro}

Driven by impressive experimental advances in controlling and probing tailored quantum matter, projective measurements~\cite{Bakr2009,Georgescu14,Gross17,Schafer20,Altman21} have opened new avenues for probing quantum many-body systems. One of the immediate questions arising from this development is how to extract observables from projective measurement outcomes. This has led to diverse approaches, ranging from detecting antiferromagnetic order in doped Hubbard models using many-body snapshots~\cite{Mazurenko17} to extracting entanglement via randomized measurements~\cite{Brydges19}.

A related, yet conceptually distinct, question is what intrinsic structure is encoded in the probability distribution of projective measurement outcomes, and what physical information it contains. Whether phases of matter can be distinguished directly from their measurement distributions has so far been explored primarily for ground states~\cite{Mendes-Santos24,andreoni2025networktheoryclassificationquantum}. For example, recent work has shown that quantum critical points in certain spin models can exhibit a distinctive structure in the corresponding wavefunction networks, constructed by connecting spin configurations whose Hamming distance is below a characteristic cutoff~\cite{Mendes-Santos24,andreoni2025networktheoryclassificationquantum}.

In contrast, quantum dynamics typically explores states far from the ground state. A paradigmatic transition separating regimes in which quantum dynamics erases or preserves local information about the initial state is the ergodicity-breaking transition. Recent work has argued, based on wavefunction fidelity susceptibilities and adiabatic gauge potentials~\cite{pandey_claeys_20}, that ergodicity-breaking transitions can be viewed as generalizations of ground-state quantum phase transitions to highly excited states~\cite{swietek_lydzba25,tokarczyk2026}. Motivated by this connection, as well as by the successful use of snapshot analysis to characterize critical ground states, we ask whether projective measurements can provide a novel probe of nonequilibrium quantum states and their dynamics.

Here, we establish a quantitative connection between the intrinsic structure of projective measurement snapshots of nonequilibrium wavefunctions and the many-body ergodicity-breaking transition. 
To establish this connection, we consider the quantum sun model~\cite{suntajs_vidmar_2022,suntajs_deroeck_24}, a paradigmatic toy model of the many-body ergodicity-breaking transition~\cite{suntajs_vidmar_2022, suntajs_deroeck_24, swietek_hopjan_25}.

\begin{figure}[b]
    \centering
    \includegraphics[width=\linewidth]{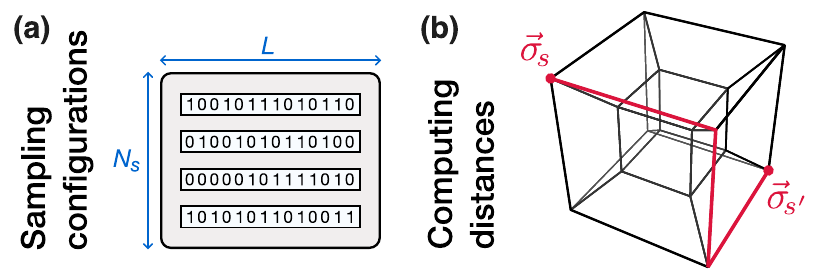}
    \caption{
    \textbf{Many-body configurations as a dataset.} \textbf{(a)} Many-body configurations of $L$ qubits,  where $L$ is the system size, are sampled from the probability distribution defined by the squared amplitudes of the wavefunction. They are collected in a dataset of size $N_s$. 
    \textbf{(b)} The metric on the data space is given by the Hamming distance $R$ between a chosen pair of many-body configurations $\vec{\sigma}_{s}$ and $\vec{\sigma}_{s'}$.
    }
    \label{fig:Schematics}
\end{figure}

We show that projective measurement snapshots of nonequilibrium wavefunctions provide two indicators of the many-body ergodicity-breaking transition. 
By a “snapshot,” we mean a many-body configuration obtained from a projective measurement of the nonequilibrium wavefunction of a quantum system comprising $L$ coupled qubits, or spins-1/2. 
We stochastically generate $N_s$ such snapshots, which form the dataset [Fig.~\ref{fig:Schematics}(a)], and characterize their intrinsic structure by using the Hamming distance between configurations [Fig.~\ref{fig:Schematics}(b)]. 
Crucially, the number of snapshots in the dataset is much smaller than the many-body Hilbert-space dimension, $N_s\ll{\mathcal D}$.

We first characterize many-body ergodicity breaking using the intrinsic dimension (ID) of the sampled dataset - a concept originally devised in the context of high-dimensional statistics and data science, which quantifies the effective number of degrees of freedom required to describe a given dataset~\cite{Campadelli2015}. In the context of many-body lattice spin models, the ID provides a natural measure of how extensively the measurement configurations explore the many-body configuration space~\cite{mendes2021unsupervised}. Indeed, in the ergodic phase, the wavefunction spreads extensively across the many-body configuration space, and we therefore expect the sampled configurations to exhibit a large - in fact, maximal - ID. In contrast, localization confines the wavefunction to a restricted region of configuration space, resulting in a smaller ID. This suggests the ID as a natural and quantitative diagnostic of the ergodicity-breaking transitions. However, conventional ID estimators suffer from the curse of dimensionality~\cite{Campadelli2015}: by inferring the ID locally from neighborhoods around individual data points, they require an exponentially large number of samples for accurate estimation, leading to a systematic underestimation of the ID when the number of features grows. 

Here, we employ the recently introduced binary intrinsic dimension (BID)~\cite{acevedo2025unsupervised,Erazo2026}, which overcomes these limitations by estimating ID globally from the distribution of pairwise distances across the entire dataset, and is particularly well-suited to many-body quantum systems with binary variables. Applied to the quantum sun model, the BID accurately identifies the ergodicity-breaking transition and exhibits qualitatively distinct behavior in the three dynamical regimes. In the ergodic phase, the BID grows in time and eventually saturates at the Thouless time to its maximal value. At the transition, the growth becomes submaximal, indicating that the measurement configurations explore an extensive but lower-dimensional structure. In the nonergodic phase, by contrast, the BID saturates at a value that is independent of system size.

We complement the intrinsic-dimension analysis by characterizing correlations within the measurement dataset through wavefunction networks~\cite{Mendes-Santos24,andreoni2025networktheoryclassificationquantum}. 
The network is constructed by connecting pairs of measurement configurations, such as $\vec{\sigma}_{s}$ and $\vec{\sigma}_{s'}$ in Fig.~\ref{fig:Schematics}(b), whenever their Hamming distance falls below a cutoff.
The distinct structures of the measurement dataset across the ergodic, critical, and nonergodic regimes are reflected in the corresponding network topology. 
In the ergodic phase, networks are homogeneous and well described by Erdős–Rényi random graphs, whereas they become increasingly heterogeneous as they approach the ergodicity-breaking transition. At the critical point, networks exhibit scale-free connectivity over several decades, with the emergence of highly connected nodes, or hubs. 
Such networks share similarities with those that appear unrelated to quantum many-body physics, such as citation networks, the World Wide Web, and certain biological and ecological systems~\cite{barabasi2016network}.
In the nonergodic phase, by contrast, a relatively small cutoff distance is sufficient to generate networks containing large fully connected components. This connectivity condensation is consistent with the small, system-size-independent BID in the nonergodic phase, revealing a complementary manifestation of the reduced dimensionality of the sampled data space.

The paper is organized as follows.
In Sec.$~$\ref{sec:QSM} we discuss the modeling of the many-body ergodicity breaking transition via the quantum sun model. 
We then study the BID of nonequilibrium quantum states in Sec.~\ref{sec:BID}, focusing on the time evolution from the initial product states as well as for the mid-spectrum eigenstates.
The wavefunction networks of these nonequilibrium states are constructed in Sec.~\ref{sec:WFN}, where we quantitatively study their degree distributions.
We conclude in Sec.$~$\ref{sec:conclusion}.

\section{Quantum Sun Model and ergodic-to-nonergodic transition}
\label{sec:QSM}

A paradigmatic model of many-body ergodicity breaking is the quantum sun model~\cite{suntajs_vidmar_2022, suntajs_deroeck_24}, which is a qubit-based model composed of an ergodic dot with $N$ qubits, coupled to $L'=L-N$ qubits outside the dot.
The model originates from studies of the avalanche mechanism of thermalization in disordered spin-1/2 chains~\cite{deroeck_huveneers_17, Luitz_17}.
It is understood that the model undergoes an ergodic to non-ergodic transition at the critical value of the tuning parameter $\alpha = \alpha_c$. 
The existence of the ergodicity breaking transition has been established using both theoretical arguments~\cite{deroeck_huveneers_17, Luitz_17, suntajs_vidmar_2022, kliczkowski_vidmar2024} as well as numerical studies of distinct transition indicators, e.g., short-range~\cite{Luitz_17, suntajs_vidmar_2022, suntajs_deroeck_24, kliczkowski_vidmar2024, pawlik_sierant_2024} and long-range~\cite{suntajs_vidmar_2022} spectral correlations, eigenstate properties such as participation entropies~\cite{suntajs_deroeck_24} and entanglement entropies~\cite{suntajs_deroeck_24, swietek_hopjan_25}, or sensitivity of eigenstates to perturbations~\cite{swietek_lydzba25}.

In our calculations, we consider the quantum sun model with $U(1)$ symmetry, focusing on the zero magnetization sector with the Hilbert-space dimension $\mathcal{D}=\binom{L}{L/2}$.
The corresponding Hamiltonian is
\begin{equation}
\label{eq:qsun:U1}
    H =H_{\mathrm{dot}}^{U(1)} + \sum_{j=0}^{L'-1} h_j\, S^z_j
     +g_0\sum_{j=0}^{L'-1} \alpha^{u_j}\, (S^x_{n(j)} S^x_j + S^y_{n(j)} S^y_j)\;,
\end{equation}
where $g_0 = 1$.
The two sums in Eq.~\eqref{eq:qsun:U1} run over $L'$ qubits outside the ergodic dot, and $n(j)$ in the last sum denotes a randomly chosen qubit inside the dot to which the outer qubit (with index $j$) is coupled to.
The distance of the outer qubits to the ergodic dot is $u_j$, which is drawn uniformly from the interval $[j - \zeta_j,\, j + \zeta_j]$, with $\zeta_j=0.2$ for $j>0$, and $u_j=0$ for $j=0$. 
In addition, each outer qubit experiences a random magnetic field $h_j$, drawn uniformly from the range $[1 - W,\, 1 + W]$, where $W=0.5$. 

The operator $\hat H_{\mathrm{dot}}^{U(1)}$ in Eq.~\eqref{eq:qsun:U1} is a random-matrix-based operator that acts non-trivially only on the $N=3$ qubits within the ergodic dot, and it satisfies the $U(1)$ symmetry. 
Here, we use a slightly different implementation of this operator than in Ref.~\cite{pawlik_zakrzewski_2024}, ensuring it has a unit Hilbert-Schmidt norm.
Details about the numerical implementation of the model in Eq.~\eqref{eq:qsun:U1} are given in Appendix~\ref{app:QSM}.

The ergodicity-breaking transition in the model is governed by the coupling $\alpha$ between the inner and outer qubits.
In Appendix~\ref{app2}, we present the numerical results for the spectral and eigenstate analysis to determine the transition point at $\alpha = \alpha_c \approx 0.75$.
This value is used as the ergodicity-breaking transition point in the remainder of the paper.

\section{Stochastic spreading in many-body Hilbert spaces: Binary intrinsic dimension}
\label{sec:BID}

We now present our first data-driven approach to directly characterize ergodicity-breaking transitions in the quantum sun model from the datasets of wavefunction snapshots, i.e., the intrinsic dimension (ID).
It is defined as follows. Let us consider a set of data points that live in some high-dimensional data space. The correlations in the data can constrain it to lie on a lower-dimensional manifold, allowing us to describe it utilizing a number of variables that is smaller than the maximum allowed, at the price of negligible information loss. The dimension of said manifold is exactly the ID: it represents the minimum number of coordinates needed to describe the data manifold. Estimating ID is not a trivial task, and several methods have been developed in the data-science literature~\cite{Campadelli2015} and applied to statistical physics phenomena~\cite{mendes2021unsupervised, mendes2021intrinsic, schmitt2022observations, verdel2024data, verdel2025family, cao2024unsupervised, Erazo2026}.

For a quantum many-body system of qubits, or spins-$1/2$, such as the quantum sun model considered here, datasets are obtained by sampling qubit configurations from a quantum many-body state. The sampling consists of measuring the state in the computational basis, or, in other words, drawing configurations in the computational basis according to Born's rule. In the resulting dataset, hence, each data point is a wavefunction snapshot, represented by a binary string, see Fig.~\ref{fig:Schematics}(a). This motivates the use of a specialized intrinsic-dimension estimator~\cite{acevedo2025unsupervised}, namely, the binary intrinsic dimension (BID). BID infers the intrinsic dimension from the global distribution of pairwise distances, making it less susceptible to the curse of dimensionality, which often causes other local estimators to underestimate the intrinsic dimension.

\begin{figure}
    \centering
    \includegraphics[width=\linewidth]{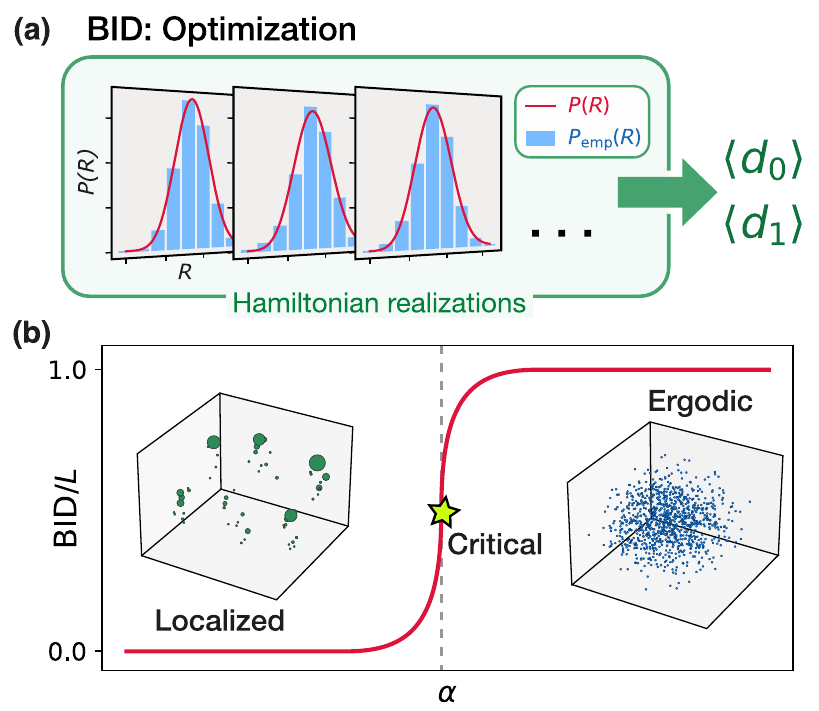}
    \caption{
    \textbf{Schematic description of BID as a probe of ergodicity breaking.} 
    \textbf{(a)} Computation of BID: 
    Given a value of the control parameter $\alpha$ and a single Hamiltonian realization, after sampling configurations and computing the Hamming distances among all possible pairs, as shown in Fig.~\ref{fig:Schematics}, the Hamming distances' empirical distribution $P_\mathrm{emp}$ is computed and interpolated with the ansatz $P(R)$ in Eq.~\eqref{eq:BID_ansatz} to obtain the parameters $d_{0}$ and $d_{1}$. For each value of the control parameter $\alpha$, the procedure is performed for many Hamiltonian realizations, and the results are averaged to compute $\langle d_{0}\rangle$ and $\langle d_{1}\rangle$, as explained in Sec.~\ref{sec:bid_tevo}.
    \textbf{(b)} Ideal ID versus the control parameter $\alpha$, for large $L$. In the ergodic phase, $\mathrm{BID}/L\simeq1$, while in the localized phase, $\mathrm{BID}/L\simeq0$. In the specific analyses of Secs.~\ref{sec:bid_latetime} and~\ref{sec:bid_eigen}, $\mathrm{BID}$ will represent the averaged quantities defined contextually. Left and right insets show 3-dimensional PCA projections~\cite{PCA_footnote,Mehta2019} of the configurations sampled from the quantum sun model with $L=18$ deep in the two regimes, respectively, at $\alpha=0.5$ and $\alpha=0.94$.
    }
    \label{fig:BIDgeneric}
\end{figure}

Let us represent the qubit configurations as points living on the $2^L$ vertices of an $L$-dimensional hypercube, $L$ being the number of qubits, as represented in Fig.~\ref{fig:Schematics}(b). This data-space can be endowed with a metric, the most natural one for binary variables being the Hamming distance,
\begin{equation} \label{def_R_Hamming}
    R(\vec{\sigma}_{s}, \vec{\sigma}_{s'}) =(L -\vec{\sigma}_{s} \cdot \vec{\sigma}_{s'})/2\,,
\end{equation}
where $\vec{\sigma}_s = \{\sigma_{1,s},\dots,\sigma_{L,s}\}$ is a sampled configuration, $\sigma_{i,s}$ is a binary variable, and $s=1,\dots,N_s$. 
If these configurations are uncorrelated, they distribute uniformly on the vertices of the hypercube, and computing the Hamming distances of all pairs of configurations would result in the following distribution:
\begin{equation}
    P_0(R) = \frac{1}{2^L} \binom{L}{R}.
\end{equation}
The presence of correlations in the datasets deforms the measured distribution $P(R)$, and the idea of the method is to reabsorb this deformation into an effective, scale-dependent dimension $d(R)$, producing the following ansatz~\cite{acevedo2025unsupervised}:
\begin{equation}
\label{eq:BID_ansatz}
    P(R) = \frac{\mathcal{C}}{2^{d(R)}} \binom{d(R)}{R},
\end{equation}
where $\mathcal{C}$ is just a normalization constant. 
Empirically, approximating $d(R)$ by retaining only the first order of the Taylor expansion, $d(R) \approx d_0 + d_1 R$, is enough to accurately reproduce the distribution. 
The parameters $d_0$ and $d_1$ are obtained by minimizing the Kullback-Leibler divergence between the empirical and the model distributions, as in Fig.~\ref{fig:BIDgeneric}(a). Consistent with the idea that the intrinsic dimension is a local property of the data manifold, the BID is defined as the small-distance limit of $d(R)$:
\begin{equation}
    \mathrm{BID} = d_0,
\end{equation}
which results in the expected $\mathrm{BID} = L$ in the uncorrelated case.

In passing, we note that the dimension of the Hilbert space $\mathcal{D}$ in the $U(1)$ symmetric quantum sun model is different from the dimension $\mathcal{D}^{\mathrm{all}}=2^{L}$ of the full binary space of $L$ qubit configurations. However, $\mathcal{D}$ still grows exponentially with $L$, i.e., $\mathcal{D}\propto 2^{L}/\sqrt{L}$, and we empirically find that the ansatz in Eq.$~$\eqref{eq:BID_ansatz} is also relevant for the  $U(1)$ symmetric the quantum sun model. 

The well-defined limits of the BID, ranging from $L$ for uncorrelated variables to $0$ for maximally correlated ones, make it a useful measure for characterizing ergodicity-breaking transitions. As illustrated schematically in Fig.~\ref{fig:BIDgeneric}(b), in the ergodic phase, high-energy eigenstates are expected to explore the available Hilbert space uniformly. Consequently, wavefunction snapshots are expected to uniformly fill the configuration space, leading to an extensive BID, with $\mathrm{BID}/L\to1$. Conversely, in the localized regime, the sampled configurations become restricted to a constrained subset of the configuration space, resulting in a BID with $\mathrm{BID}/L\to0$. Analogously, the same argument can be extended for wavefunctions at late times in quench dynamics from some initial high-energy product state.

In the following, we test these predictions by analyzing the BID across the ergodicity-breaking transition in the quantum sun model. We first investigate the dynamical evolution of the BID following a quench from a high-energy initial state, focusing on its growth and late-time saturation, which is particularly relevant to experiments. We then complement this with the analysis of the BID for the mid-spectrum eigenstates. Our analysis will focus explicitly on the basis compatible with the $U(1)$ symmetry; however, based on what has been found in the context of low-energy physics~\cite{andreoni2025networktheoryclassificationquantum}, we expect similar results for other basis choices as well.

\subsection{Time evolution of BID}
\label{sec:bid_tevo}

\begin{figure*}
    \centering
    \includegraphics[width=\linewidth]{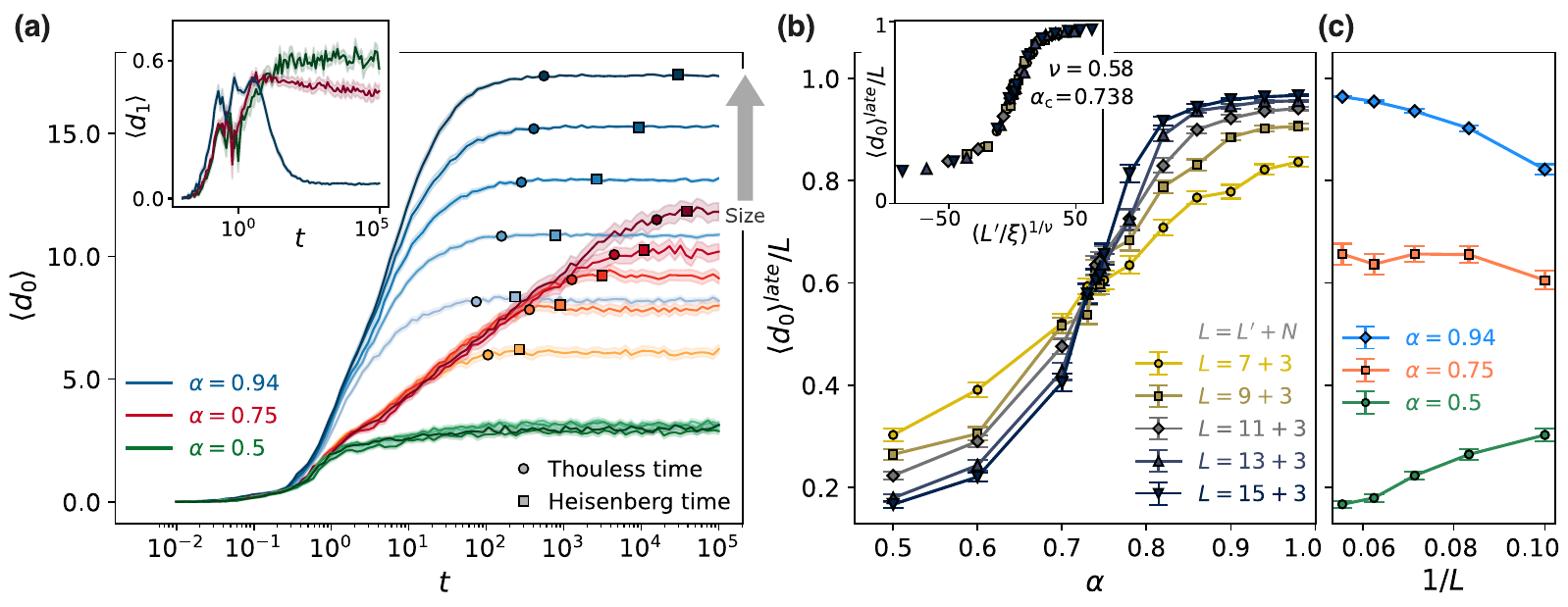}
    \caption{\textbf{Time evolution of BID. (a)}
    BID $\langle d_0 \rangle$ versus time for $\alpha=0.5, 0.75, 0.94$, at different system sizes $L=10-18$, represented by lighter to darker shades of each color.  Solid lines and shaded areas around them are averages at fixed times $t$ and standard deviations of the BID over 100 Hamiltonian realizations;
    for each of them, the BID was computed from a dataset of $N_s=100$ snapshots, sampled with Born's rule from the time-evolved wavefunction $|\Psi(t)\rangle$. Circles and squares represent Thouless and Heisenberg times, respectively. The inset represents the disorder-averaged values of $\langle d_1 \rangle$ as a function of time, for the same values of $\alpha$, at $L=18$.
    \textbf{(b) }Late-time values of the BID versus the interaction strength $\alpha$, for different system sizes. Values and error bars are computed as the averages and standard deviations of the means of the single-time values after reaching the plateau at late times. At each time $t$, BID has been computed from a dataset of $N_s=100$ configurations sampled from the time-evolved wavefunction $|\Psi(t)\rangle$ according to Born's rule. The inset shows the collapse. To estimate $\alpha_c$, we perform the scaling collapse of finite-size results with an ansatz as in Ref.~\cite{swietek_hopjan_25}, i.e., we consider the correlation length $\xi=[\ln(\alpha/\alpha_c)^2]^{-\nu}$, where $\nu$ is the critical exponent. 
    \textbf{(c) }Same values as in panel (b), for $\alpha=0.5,0.75,0.94$, as a function of the inverse system size, to show the approach to the values expected in the infinite-size limit, i.e., $\mathrm{\rm BID}/L\to0,1$ in non-ergodic and ergodic phases, respectively, and a finite value between 0 and 1 at the critical point. }
    \label{fig:BID_timeEVO}
\end{figure*}

To study the dynamics of BID, we start with an initial product state in the computational basis. To probe the properties of highly excited states and avoid finite-size effects, we choose the initial state $|\psi(0)\rangle$ as the product state with the energy closest to the average energy, i.e., $\langle \psi|H|\psi\rangle \approx {\rm Tr}(H)/\mathcal{D}$. We perform unitary time evolution under the quantum sun model Hamiltonian in Eq.~\eqref{eq:qsun:U1}, $|\psi(t) \rangle=e^{-iHt}|\psi(0)\rangle$. At each time $t$, we sample $N_s$ configurations from the time-evolved state, and compute $d_0$ (BID) and $d_1$ of the sampled dataset. Finally, we average both the $d_0$ and $d_1$ over $100$ different Hamiltonian realizations and denote the average quantities as $\langle\,\cdot\,\rangle$.

In the main panel of Fig.$~$\ref{fig:BID_timeEVO}(a), we show the time evolution of the averaged BID $\langle d_0\rangle$. We choose the parameters in the ergodic and non-ergodic phases of the quantum sun model, represented by $\alpha=0.94$ and $\alpha=0.5$, respectively, as well as the critical point at $\alpha_c \approx 0.75$. 

We first examine the growth of BID in the ergodic phase ($\alpha=0.94$) and present the results in the main panel of Fig.$~$\ref{fig:BID_timeEVO}(a), see the blue lines, for different system sizes $L= 10-18$. We observe that the BID grows in time and saturates to values that are close to the maximal values, $d^\mathrm{max}_0 = L$, consistent with the complete delocalization of the initial state. 
Importantly, the saturation of $\langle d_0 \rangle$ occurs at the Thouless time $t_{\rm Th}$, which is fast compared to the Heisenberg time $t_{\rm H}$, shown as circles and squares on the blue curves, respectively. The Heisenberg time is defined as the inverse of average level spacing, i.e., as $2\pi/\langle \delta E_\mu \rangle_\mu $, where $\delta E_\mu = E_{\mu+1} - E_\mu$ is the level spacing between the eigenlevels $\mu$ and $\mu-1$, respectively, and $\langle \cdots \rangle_\mu$ denotes the average over all eigenstates. We further average the Heisenberg time over Hamiltonian realizations to obtain $t_{\rm H}=\langle2\pi/\langle \delta E_\mu \rangle_\mu \rangle_H$. The Thouless time is obtained from the spectral analyzes as the onset of universal ergodic behavior; see Appendix~\ref{app2}. 
Physically, the Heisenberg time sets the upper bound for the accessible time scales in the finite system, and the Thouless time is usually interpreted as the longest physically relevant time scale. 
Here, the Thouless time can be interpreted as the saturation time of information spreading in Hilbert space, which recalls its interpretation in short-range quadratic systems, i.e., the saturation time of the mean-square displacement~\cite{hopjan_vidmar_25}.

The physical interpretation of these results is clear: the discrete manifold spanned by snapshots is indicative of the system capability of exploring the full configuration space - even if it is heavily undersampled. Moreover, the timescale at which such saturation occurs is directly tied to the Thouless time, which is a clear hint of the physical significance of BID.

We next consider the critical point ($\alpha=0.75$) and present the dynamics of BID in the main panel of Fig.$~$\ref{fig:BID_timeEVO}(a), represented by the red lines therein. The BID grows logarithmically in time and saturates to a value that depends on the system size. Importantly, the saturation value of the BID is not maximal. The observed sub-maximal saturation indicates that the data structure occupies a lower-dimensional manifold in configuration space.
Note that the proportionality between the Thouless and the Heisenberg time
is clearly visible, as the critical dynamics emerge when the scalings of the Thouless and Heisenberg times approach each other~\cite{suntajs_bonca_20a, sierant_delande_20, suntajs_vidmar_2022, hopjan_vidmar_23, hopjan_vidmar_25}. 
Indeed, the value of the critical point can be estimated from the condition $t_\mathrm{Th}/t_{\rm H}=\mathrm{const}$, see Fig.~\ref{fig:qs_gap_ratio} in Appendix~\ref{app2} for details. 

A notable consequence of the criticality
defined via the condition $t_{\rm Th}(L) \propto t_{\rm H}^{}(L)$ is the emergence of the scale invariant principle, similar to that for the survival probability~\cite{hopjan_vidmar_23, hopjan_vidmar_23b, hopjan_vidmar_25}. Indeed, a subtracted BID by its late time value, i.e., $\langle \tilde{d}_0(t) \rangle=\langle d_0(t)\rangle-\langle d_0 \rangle^\mathrm{late}$, becomes scale-invariant if time is measured in units of the Heisenberg time, $t \rightarrow \tau=t/t_{\rm H}$ (not shown). Such scale-invariant, or self-similar, behavior of $\langle \tilde{d}_0(\tau) \rangle$ is another manifestation of the critical dynamics.

Finally, in the non-ergodic phase ($\alpha=0.5$), the BID curves for different system sizes overlap, see the green lines in Fig.~\ref{fig:BID_timeEVO}(a). 
This suggests saturation to a system-size independent value that can be interpreted as localization in the Hilbert space.

For all the values of $\alpha$ under consideration, i.e., $\alpha=0.5,0.75,0.94$, we also plot the time evolution of $\langle d_1 \rangle$ for the largest system size $L=18$ in the inset of Fig.$~$\ref{fig:BID_timeEVO}(a). In the ergodic phase, after reaching a maximum at mid-times, it drops at late times to a value close to zero, signaling a restoration of perfect symmetry of Hamming distance distribution in Eq.~\eqref{eq:BID_ansatz}, and a uniformity in the distribution of the sampled configurations in configuration space. The saturation of $\langle d_1 \rangle$ to non-zero values in the other phases suggests that the distributions in Eq.~\eqref{eq:BID_ansatz} are skewed. 

\subsection{Late time values of BID}
\label{sec:bid_latetime}

As a second step in our BID analysis, we analyze the late-time saturation values of $\langle d_0 \rangle$, normalized by the system size $L$, i.e., $\langle d_0 \rangle^\mathrm{late}/L$. 
In the main panel of Fig.$~$\ref{fig:BID_timeEVO}(b), we plot $\langle d_0 \rangle^\mathrm{late}/L$ as a function of $\alpha$ for several system sizes $L$. 
In the ergodic and non-ergodic phases, i.e., at $\alpha>0.75$ and $\alpha<0.75$, respectively, the normalized BID displays two opposite flows towards the limiting values $\langle d_0\rangle^\mathrm{max}/L=1$ and $\langle d_0\rangle^\mathrm{min}/L=0$, with a crossing at the critical point. 
The flow with inverse system size can be better appreciated in Fig.$~$\ref{fig:BID_timeEVO}(c) for the representative points $\alpha=0.94$, $\alpha=0.5$ and $\alpha=0.75$. In particular, one can observe a saturation around a sub-maximal value $\langle d_0\rangle^\mathrm{late}/L\simeq0.65$ at the critical point.
This behavior of $\langle d_0 \rangle^\mathrm{late}/L$ resembles that of the short-range spectral statistics measured by the gap ratio measure, which flow to the Gaussian Orthogonal Ensemble (GOE) and Poisson limit in the ergodic and non-ergodic phases, respectively, and exhibit intermediate statistics at the ergodicity-breaking transition, see Fig.~\ref{fig:qs_gap_ratio} in Appendix~\ref{app2}. 

Inspired by this similarity, we perform a finite-size scaling analysis of the BID data, following the ansatz of Ref.~\cite{swietek_hopjan_25}: we consider the correlation length $\xi=[\ln(\alpha/\alpha_c)^2]^{-\nu}$, where the critical exponent $\nu$ and $\alpha_c$ are the fitting parameters. We plot the results as a function of $(L'/\xi)^{1/\nu}$ and perform the optimization of the data collapse via the cost function minimization from Ref.~\cite{suntajs_bonca_20b}. Remarkably, we obtain $\alpha_c^\mathrm{BID}\approx 0.738$ which is very close to the value estimated from the gap ratio statistics, $\alpha_c^{r}\approx 0.749$, compare collapses in the insets of Fig.$~$\ref{fig:BID_timeEVO}(b) and Fig.~\ref{fig:qs_gap_ratio}(a) in Appendix~\ref{app2}. 
As a technical remark, we note that $\nu$ is expected to flow to its limiting value $\nu=1$, for large systems close to the phase transition~\cite{swietek_hopjan_25}. In the finite size numerical analyses, it usually has smaller values~\cite{swietek_hopjan_25}; the value $\nu^\mathrm{BID}=0.575$ obtained here is aligned with this expectation. 

Hence, it is remarkable that the normalized BID accurately identifies the position of the critical point, which is in close agreement with predictions from conventional spectral methods.
This occurs despite the BID being calculated from quantum dynamics at finite times, and using only a moderate number of snapshots, $N_s\ll\mathcal{D}$. 

\begin{figure}[t!]
    \centering
    \includegraphics[width=\linewidth]{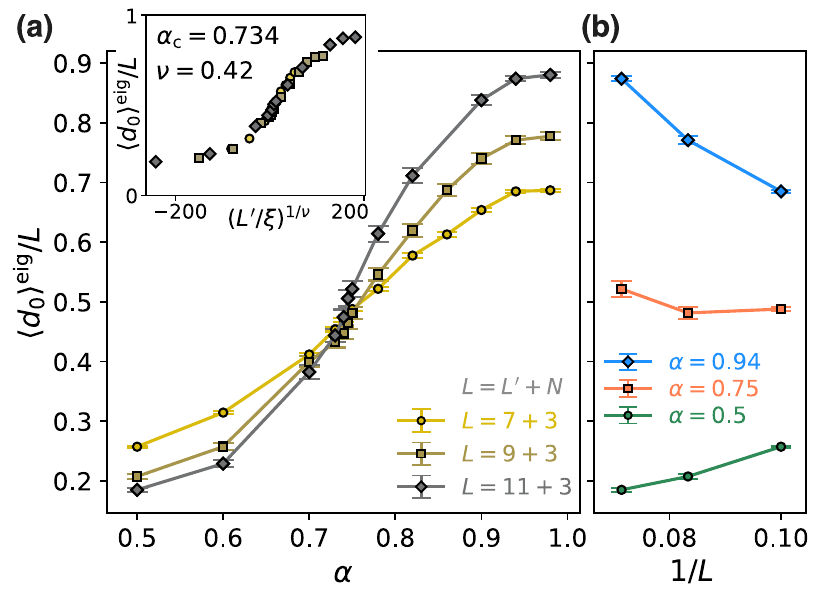}
    \caption{\textbf{Mid-spectrum BID. } \textbf{(a) }BID versus interaction strength $\alpha$ for different system sizes $L$, computed from the mid-spectrum eigenstate analysis. Values and errorbars are computed as the average and standard deviation of the BID computed by sampling the Hamiltonian's eigenstates in the middle of the spectrum, for 1000 realizations of the disorder. We consider 25 Hamiltonian eigenstates for $L=7+3$, and 500 for all other system sizes. For each combination of eigenstate and Hamiltonian realization, BID has been computed from a dataset of $N_s=100$ samples from the eigenstate's wavefunction according to Born's rule. Inset shows the collapse. To estimate $\alpha_c$, we perform scaling collapse of the results with an ansatz as in Ref.~\cite{swietek_hopjan_25}, i.e., we consider the correlation length $\xi=[\ln(\alpha/\alpha_c)^2]^{-\nu}$, where $\nu$ is the critical exponent. 
    \textbf{(b) }Same values as panel (a), for $\alpha=0.5,0.75,0.94$, as a function of the inverse system size, to show the approach to the values expected in the infinite-size limit, i.e., $\mathrm{\rm BID}/L\to0,1$ in non-ergodic and ergodic phases, respectively, and a finite value between 0 and 1 at the critical point.
    }
    \label{fig:BID_eigen}
\end{figure}

\subsection{BID of the mid-spectrum eigenstates}
\label{sec:bid_eigen}

We complement our analysis of the BID dynamics with the study of the BID of highly excited eigenstates in the middle of the many-body energy spectrum. To this end, we perform exact diagonalization and collect up to $500$ eigenstates in the middle spectrum. For each eigenstate, we consider $100$ wavefunction snapshots to compute the BID, and finally, we average the results over $1000$ Hamiltonian realizations. 

We consider the averaged eigenstate BID denoted by $\langle d_0\rangle^{\rm eig}$, rescaled by the system size $L$, i.e., $\langle d_0 \rangle^\mathrm{eig}/L$. 
In Fig.~\ref{fig:BID_eigen}(a), we plot $\langle d_0 \rangle^\mathrm{eig}/L$ versus $\alpha$ for several system sizes $L$, and in Fig.~\ref{fig:BID_eigen}(b) we plot $\langle d_0 \rangle^\mathrm{eig}/L$ versus the inverse system size.
This study is analogous to the one of the late-time BID, $\langle d_0 \rangle^\mathrm{late}/L$.
Remarkably, results for the eigenstate BID in Fig.~\ref{fig:BID_eigen}, and results for the late-time BID in Figs.~\ref{fig:BID_timeEVO}(b) and~\ref{fig:BID_timeEVO}(c), share many similarities.
Still, the saturation value of the eigenstate BID at the critical point slightly differs from that of the late-time BID; compare $\langle d_0\rangle^\mathrm{eig}/L\simeq 0.5$ in the Fig.$~$\ref{fig:BID_eigen}(a)
to $\langle d_0\rangle^\mathrm{late}/L\simeq 0.65$ in Fig.$~$\ref{fig:BID_timeEVO}(b).

Finally, we perform the finite-size scaling analysis of the eigenstate BID, see the collapse in the inset of Fig.$~$\ref{fig:BID_eigen}(a), analogous to the one done for the late-time BID, see the inset of Fig.$~$\ref{fig:BID_timeEVO}(b). 
We obtain $\alpha_c^\mathrm{BID,eig}\approx 0.737$ and  $\nu^\mathrm{BID,eig}=0.42$ for the eigenstate BID, which are in good agreement with $\alpha_c^\mathrm{BID}\approx 0.738$ and $\nu^\mathrm{BID}=0.58$ for the late-time BID.
We remark that such values have to be taken {\it cum grano salis}, as we have used only three system sizes in the eigenstate BID analysis.

\section{Correlations beyond dimensionality: wave-function networks and multifractality}
\label{sec:WFN}

\subsection{Network construction}

We next go beyond dimensional reduction by assessing how the features of ergodicity-breaking transitions reflect in the datasets' correlations. To this end, we focus on the wavefunction networks~\cite{Mendes-Santos24}, i.e., networks constructed from the given datasets, and analyze the properties of these networks. 

From a dataset of wavefunction snapshots, the network is constructed as follows. We first associate each sampled configuration with a node of the wavefunction network. Second, we compute the distances between the nodes of the network using a chosen metric. In our case, the natural choice for this metric is the Hamming distance $R$ in Eq.~\eqref{def_R_Hamming} between two configurations, which is also directly related to Parisi two-replica overlap. The last step along the construction is the activation of links between nodes. We activate links between pairs of data points whose distance is below or equal to a chosen cutoff, i.e., $R\leq R^{*}$. The resulting network is thus geometrical in nature.

While the choice of the cutoff appears arbitrary, we argue that the main conclusions are independent of it. In fact, there are two natural choices for the cutoff radius $R^{*}$. The first is the average distance of the $n$-th nearest neighbor, denoted as $R^{*}_n$, which allows the network to choose the cutoff by itself as a natural locality-induced distance in the dataset. The other choice is to set $R^{*}$ to be an even integer. We first focus on the former, which was previously used in the study of ground-state wavefunctions~\cite{Mendes-Santos24, andreoni2025networktheoryclassificationquantum}, and then connect it to the latter. 

The average distance (over the dataset) of the $n$-th neighbor is defined as~\cite{Mendes-Santos24,andreoni2025networktheoryclassificationquantum}: 
\begin{equation}\label{eq:cutoff_def}
    R^{*}_n =
    \frac{1}{N_s} \sum_{s=1}^{N_s} R(\vec{\sigma}_{s}, \vec{\sigma}_{s'}) \,\Big|_{{s'}=n^\text{th} \mathrm{\;n.n.\;of\;}{s}},     
\end{equation} 
where, for a fixed $s$, the Hamming distance $R(\vec{\sigma}_{s}, \vec{\sigma}_{s'})$ is calculated between the configuration $\vec{\sigma}_{s}$ and the configuration $\vec{\sigma}_{s'}$, which corresponds to the $n^\text{th}$ nearest neighbor relative to $\vec{\sigma}_{s}$. 
The ordering of neighbors is induced by the Hamming distance, from the smallest to the largest, and the nodes of equal distance are sorted into a single group. 
The ordering of the nodes within a group is irrelevant when computing the averaged cutoff $R^{*}_n$. 

\subsection{Network structure and degree distribution}

Once the network is constructed, we study the distributions $P(k)$ of the degree $k$ of the network nodes, dubbed the degree distribution.
It is defined as $P(k) = N_k/N_s$, where $N_k$ is the number of nodes with degree $k$.

\begin{figure}
    \centering
    \includegraphics[width=\linewidth]{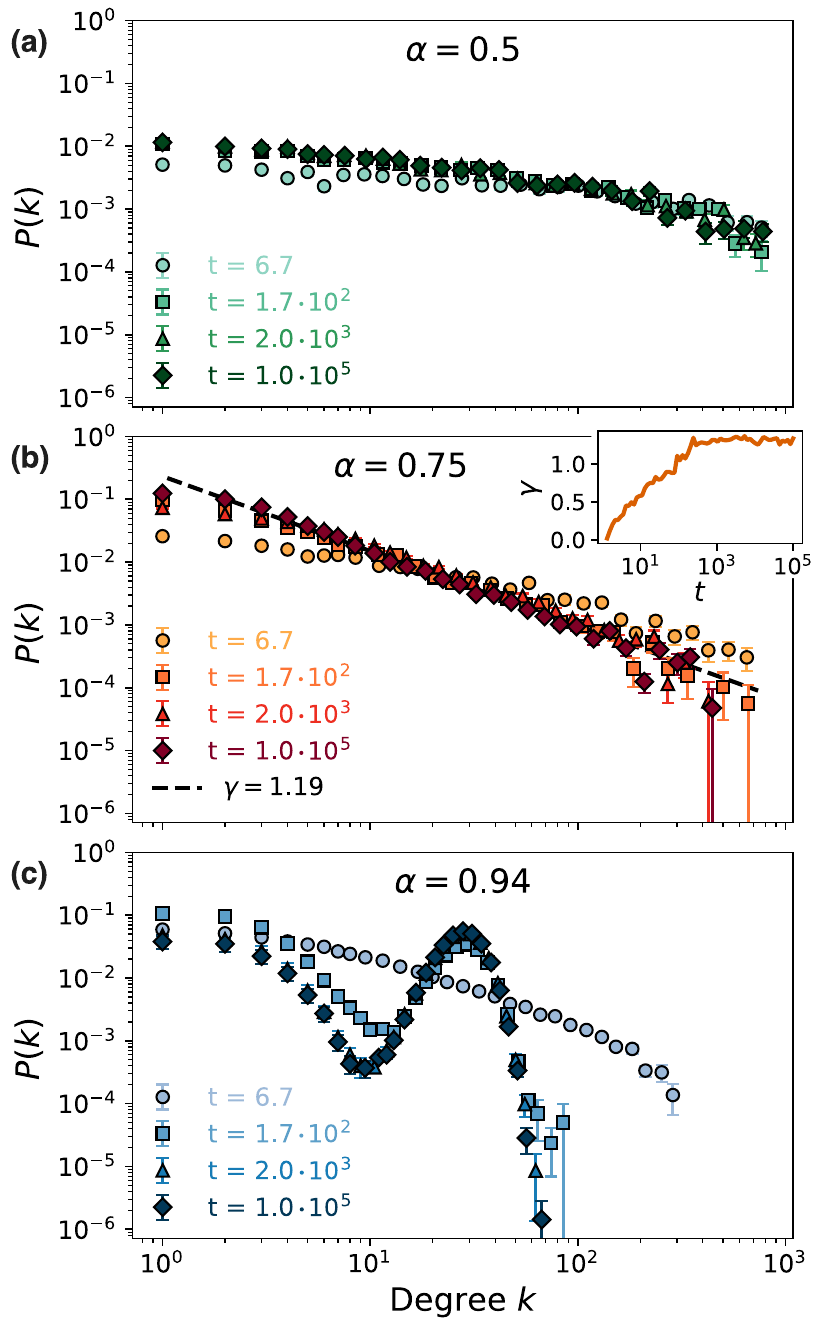}
    \caption{\textbf{Time evolution of wavefunction networks.} \textbf{(a), (b), (c)} Degree distributions $P(k)$ of the averaged wavefunction networks, built from configurations sampled for $L=18$ at selected times in the non-ergodic phase ($\alpha=0.5$), at the critical point ($\alpha=0.75$), and in the ergodic phase ($\alpha= 0.94$), respectively. The averaged degree distributions are obtained by averaging over 100 distributions, each of them obtained separately for different Hamiltonian realizations, with $N_s=1000$ and the cutoff $R^{*}_n$, where $n=10$. The dashed line in panel (b) represents the fit with a power law $P(k)\propto k^{-\gamma}$ for $k\in[8, 150]$. The inset shows the values of the interpolated $\gamma$ in the same region as a function of time $t$.}
    \label{fig:Networks_time}
\end{figure}

In Fig.$~$\ref{fig:Networks_time} we plot the degree distribution $P(k)$ corresponding to the non-ergodic phase, critical point, and ergodic phase, with $\alpha=0.5,\ 0.75$ and $0.94$, respectively, obtained by the snapshots at selected times. 
We set $N_s=1000$ and the cutoff radius to the average distance of the $n=10$th nearest neighbor, such that $1 \ll n \ll N_s$.
Finally, we average the degree distribution over  $100$ Hamiltonian realizations.

\begin{figure}
    \centering
    \includegraphics[width=\linewidth]{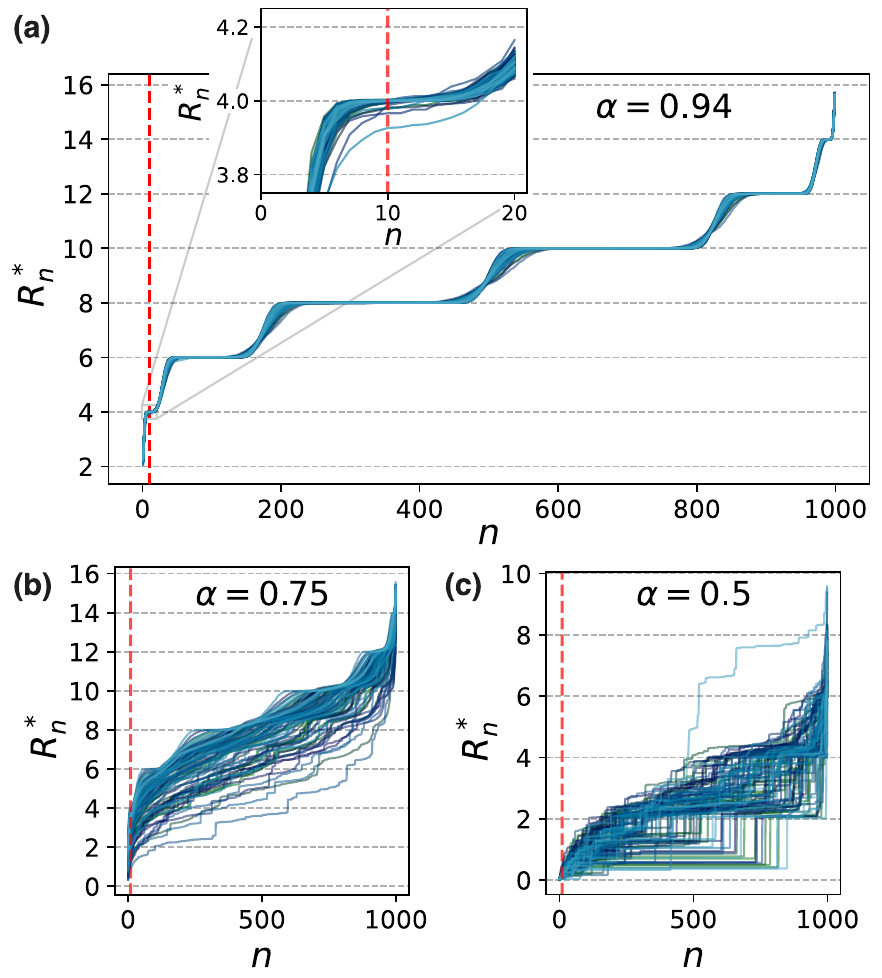}
    \caption{\textbf{The dependence of cutoff $R^{*}_n$ on $n$.} \textbf{(a)} The cutoff distance $R^{*}_n$ is computed for different values of the index $n\in[1,1000]$, for $\alpha=0.94$ in the ergodic phase, at $t=10^5$ for $L=18$. Every line is a different Hamiltonian realization, for which $R^{*}_n$ is computed from a set of $N_s=1000$ sampled configurations. Vertical dashed lines denote the value $n=10$ chosen for the distributions in Fig.$~$\ref{fig:Networks_time}. The inset shows the zoomed-in region around $n=10$. \textbf{(b) and (c)} Analogous plots, respectively, for the critical point, $\alpha=0.75$, and in the localized phase, $\alpha=0.5$. 
    }
    \label{fig:R_analysis}
\end{figure}

\begin{figure*}
    \centering
    \includegraphics[width=0.9\linewidth]{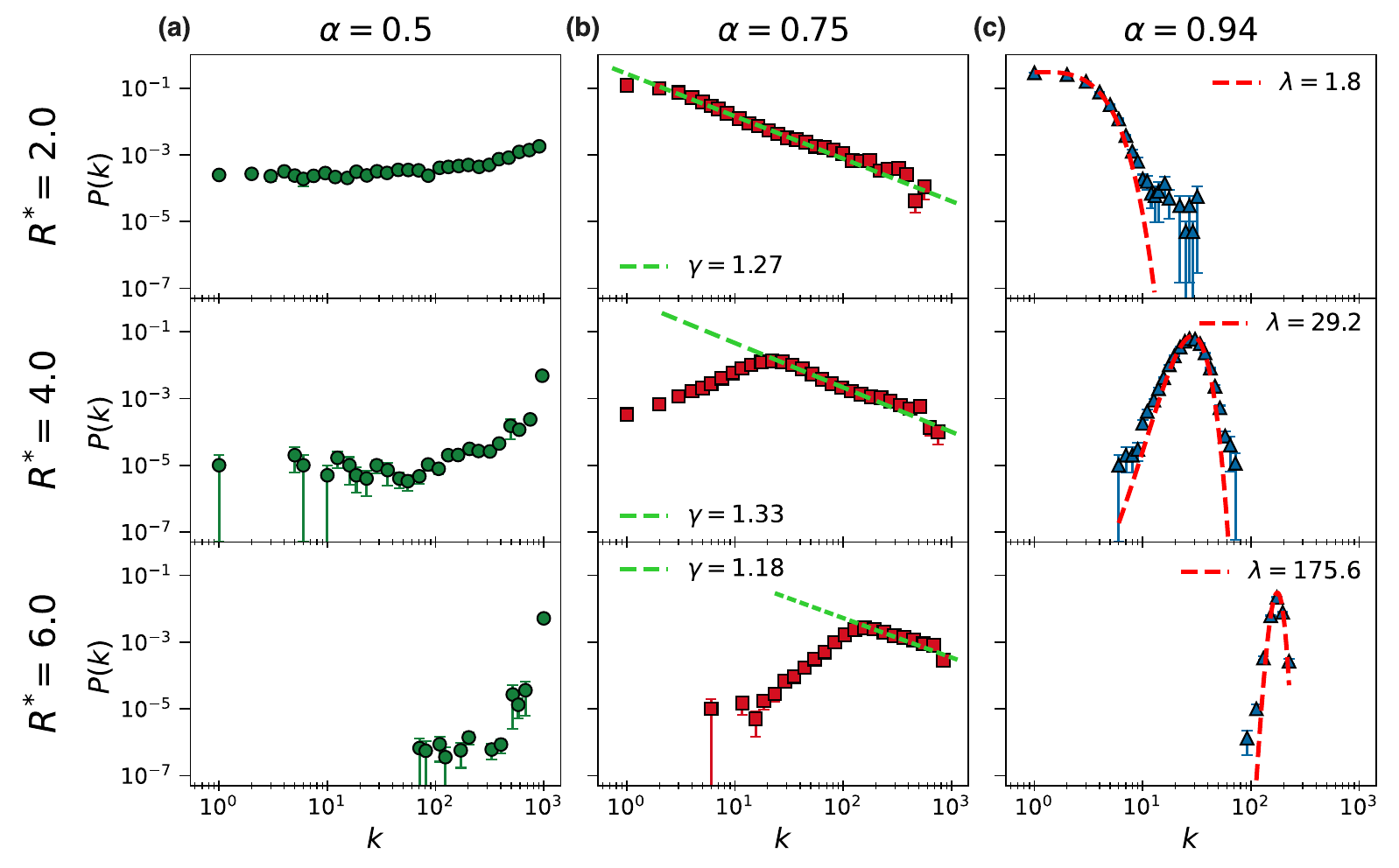}
    \caption{\textbf{Wavefunction networks with fixed cutoff-distance.} 
    Degree distributions $P(k)$ of the averaged wavefunction networks, built from configurations sampled at selected times in the non-ergodic phase ($\alpha=0.5$) {\bf (a)}, at the critical point ($\alpha=0.75$) {\bf (b)}, and in the ergodic phase ($\alpha= 0.94$) {\bf (c)}, respectively. The averaged degree distributions are obtained by averaging over 100 distributions, each of them obtained separately for different Hamiltonian realizations, with $N_s=1000$. The cutoff $R^{*}$ is chosen to be a different fixed integer value for the top, middle, and bottom row, respectively $R^{*}=2,4,6$. 
    The dashed green lines in panel (b) represent the fit with a power law $P(k)\propto k^{-\gamma}$ for $k\in[5, 100]$, $k\in[20, 300]$, and $k\in[150, 700]$, respectively, in the top, middle, and bottom row. The fits resulted in $\gamma \approx 1.27, 1.39, 1.18$, respectively.
    The dashed red lines in panel (c) represent the fit with a Poissonian $P(k) \propto \lambda^k e^{-\lambda}/k!$. The fits resulted in $\lambda \approx 1.8, 29.2, 175.6$, respectively.
    }
    \label{fig:Networks_fixR}
\end{figure*}

We observe distinct features of the distributions across the ergodicity-breaking transition. 
In the non-ergodic phase, see Fig.~\ref{fig:Networks_time}(a), $P(k)$ is approximately constant, with a slight tilt. 
In the ergodic phase, see Fig.~\ref{fig:Networks_time}(c), $P(k)$ develops a double peak structure with exponential decays of $P(k)$ from the centers of the peaks. 
Below, we inspect the functional forms of $P(k)$ in both phases in more detail.
We find that, while they both exhibit the characteristic network structures that are unique to the underlying phases, certain features, such as the double peak structure in Fig.~\ref{fig:Networks_time}(c), are consequences of the specific choice of the cutoff $R_n^*$.

Remarkably, at the ergodicity-breaking transition point, see Fig.~\ref{fig:Networks_time}(b), the degree distribution $P(k)$ exhibits a power-law decay,
\begin{equation}\label{eq:sf_networks}
    P(k) \propto k^{-\gamma}\;,
\end{equation} 
with an exponent $\gamma \approx 1.19$. 
This is the main result of this section, suggesting that the wavefunction networks at the critical point are scale free on several orders of magnitude.
Below, we also argue that the value of the exponent $\gamma$ of the power-law distribution is, to a large extent, independent of the specific choice of the cutoff $R_n^*$.

We note that the wavefunction networks are constructed from nonequilibrium wavefunctions, and hence they depend on time.
While we are in principle interested in properties at late times, we also observe that the
power-law decay of $P(k)$ develops at times that are much shorter than the Thouless time $t_{\rm Th}$.
This is shown in the inset of Fig.$~$\ref{fig:Networks_time}(b), where we plot the time-resolved exponent of the power-law distribution, denoted as $\gamma(t)$.
The time when $\gamma(t)$ saturates is shorter than $t_{\rm Th}$, which is the time at which the BID saturates, see the main panel of Fig.$~$\ref{fig:BID_timeEVO}(a) at $\alpha=0.75$.

Next we study robustness of the observed functional forms of $P(k)$ in Fig.~\ref{fig:Networks_time}.
To this end, it is instructive to study the dependence of $R^{*}_n$ on $n$. 
In Fig.~\ref{fig:R_analysis}, we plot $R^{*}_n$ as a function of $n$ for each of the $100$ Hamiltonian realizations. 
In the ergodic phase, see Fig.$~$\ref{fig:R_analysis}(a), $R^{*}_n$ vs $n$ exhibits a step-like behavior where the plateaus emerge at even integers. 
In the inset of Fig.~\ref{fig:R_analysis}(a), we zoom in on the vicinity of $n=10$, and we observe that $R^{*}_n<4$ for part of the Hamiltonian realizations, and $R^{*}_n\geq 4$ for the rest of the Hamiltonian realizations. 
Due to the discreteness of the metric, the degree distribution of the networks with $R^{*}_n<4$ will hence be peaked around smaller values of $k$ when compared to those with $R^{*}_n\geq4$.
Consequently, when the degree distribution is averaged over the Hamiltonian realizations, the double peak structure emerges: the networks with $R^{*}_n<4$ and those with $R^{*}_n\geq 4$ contribute to the peaks at lower and higher values of $k$, respectively. 
Similar, even though milder effect is also present at the critical point and in the non-ergodic phase; see Figs.~\ref{fig:R_analysis}(b) and~\ref{fig:R_analysis}(c), respectively, for the profiles of $R^{*}_n$, which exhibit a weaker effect of the plateau structure, but stronger dependence on Hamiltonian realizations.
These effects may eventually hinder a clear distinction between the critical point and the ergodic and non-ergodic phases.

The above analysis suggests the need for an alternative definition of the cutoff $R^*$.
We empirically find that the optimal choice is to study the degree distributions $P(k)$ of wavefunction networks with the cutoff fixed to the even integers, i.e., $R^{*}=2,4,6$. 
In Fig.$~$\ref{fig:Networks_fixR} we plot the resulting $P(k)$, which exhibit clearly distinct behavior for all three values of $\alpha$. 

In the ergodic phase, see Fig.$~$\ref{fig:Networks_fixR}(c), the distribution $P(k)$ now exhibits a single peak and an exponential decay away from the peak.
We fit $P(k)$ with the function
\begin{equation}\label{eq:er_networks}
    P(k) \propto \lambda^k e^{-\lambda}/k! \,.
\end{equation} 
These Poisson-like profiles are associated with random Erdős–Rényi-like wavefunction networks with fixed connectivity. 
Thus, we can argue that the ergodic phase is associated with the emergence of random networks, and this finding correlates with the maximal values of the BID found in the ergodic phase.

In contrast, in the non-ergodic phase, see Fig.$~$\ref{fig:Networks_fixR}(a), we find networks with a non-zero probability for a broad range of $k$ values. For larger $R^{*}$, the networks become more and more connected, which is manifested as the accumulation of weight in $P(k)$ close to $k_\mathrm{max}=N_s-1$, see the bottom panel in  Fig.$~$\ref{fig:Networks_fixR}(a).
In the latter, most of the nodes contain $k \approx k_{\rm max}$, and hence a finite number of nodes carries a finite fraction of all connections (this is often referred to as condensed network, even if here the network has, in addition, a geometrical structure).
The origin of this behavior is the wavefunction localization in the many-body Hilbert space, as suggested by the BID analysis in Sec.~\ref{sec:BID}. 
The localization is also reflected in the many repetitions of the same data point in the dataset, which keeps $R^{*}_n<2$ until very large values of $n$, see Fig.~\ref{fig:R_analysis}(c).

Finally, at the critical point, see Fig.$~$\ref{fig:Networks_fixR}(b), the distributions $P(k)$ can be fitted to the power-law in Eq.$~$\eqref{eq:sf_networks}.
This is the case for all values of $R^*$ under investigation, provided that $k$ is high enough.
We find the values of the power-law exponent $\gamma\approx 1.27, 1.39, 1.18$ for the cutoff values $R^{*}=2,4,6$, respectively. 
Considering the sensitivity of the fits to the range of $k$ values, the values of $\gamma$ can be regarded as almost cutoff independent. The power-law dependence also appears in the averaged degree distribution with fixed $n$; compare Fig.$~$\ref{fig:Networks_fixR}(b) to Fig.$~$\ref{fig:Networks_time}(b). However, we regard the degree distributions with fixed $R^{*}=2,4,6$ as having a clearer signature of the power-law profiles. 
The power-law distributions reflect the heterogeneous structure of the networks, which contain a large number of low connected nodes, but also a few highly connected nodes called hubs. 
These networks are known as scale-free.

An intriguing property of scale-free networks is their pronounced sensitivity to targeted removal of highly connected nodes~\cite{albert_jeong_00}. 
This property may share similarities with another well-studied property of quantum states, namely, their sensitivity to perturbations. 
Indeed, ground-state quantum critical points are known to exhibit enhanced sensitivity to perturbations, as quantified, for example, by their stability under renormalization of the couplings, or by the fidelity susceptibility~\cite{quan_song_06, zanardi_giorda_07, you_gu_2007, gu_10}. 
More recently, a connection between fidelity susceptibility at arbitrary energies and the norm of adiabatic gauge potentials~\cite{pandey_claeys_20, sels_polkovnikov_21, leblond_sels_21} has established that the sensitivity to appropriately chosen perturbations is enhanced at the many-body ergodicity-breaking transition~\cite{swietek_lydzba25}.
While we are not aware of any quantitative relation between sensitivity of quantum states to perturbations, and sensitivity of wavefunction networks to removal of highly connected nodes, they both connect the structure of highly excited quantum states at the boundary of ergodicity to the structure of ground states at quantum critical points.

\section{Conclusions} 
\label{sec:conclusion}

In this work, we introduced a new perspective on characterizing nonequilibrium dynamics from stochastically generated wavefunction snapshots. Our main finding is that wavefunction snapshots allow to treat ergodicity-breaking transitions on the same footing as traditional equilibrium transitions, providing thus a unifying framework to describe such - otherwise very distinct - physical phenomena. We achieve this goal by inspecting the dynamics of information spreading in many-body Hilbert space, interpreting it as a high dimensional discrete manifold, and computing: i) its binary intrinsic dimension, and ii) the degree distributions of wavefunction networks. We show that these exhibit distinct dynamical behavior across the ergodic, critical, and non-ergodic phases. 

The dynamics of binary intrinsic dimension exhibits a change in behavior across the transition, abruptly shifting from maximal growth in the ergodic phase to saturation in the non-ergodic phase. At criticality, sub-maximal growth of the binary intrinsic dimension indicates that correlations in the data structure can be effectively modeled as an uncorrelated data set on an extensive but lower-dimensional manifold. We further show that the normalized binary intrinsic dimension accurately identifies the critical point, which is in close agreement with conventional spectral methods. 

The lack of correlation in the data structure in the ergodic phase is reflected in the emergence of random Erdős–Rényi-like wavefunction networks. In the non-ergodic phase, the localization of the wavefunction in the Hilbert space forces the wavefunction networks to become fully connected. At the critical point, correlations in the data structure, effectively captured by the lower-dimensional manifold, yield scale-free wavefunction networks with power-law tails. Remarkably, these networks can be found in many areas of science, e.g., in the structures of researcher collaboration networks and the World Wide Web.
 
For the power-law exponent $\gamma$ of scale-free networks at criticality, we get $\gamma \in [1.1,1.4]$ from different protocols of network construction.
Intriguingly, this interval overlaps with the value of the power-law exponent reported for the ground-state critical point of the quantum Ising model, $\gamma_{\rm GS}\approx 1.25$$~$\cite{andreoni2025networktheoryclassificationquantum}. 
Moreover, the emergence of scale-free networks at the critical point, in particular their heterogeneous structure with a large number of low connected nodes and a few highly connected nodes, could be associated with the multifractal character of the wave functions at the critical point.
The questions of possible universality of power-law exponents, as well as their quantitative connection to wavefunction multifractality, call for further studies in this direction.

Our study demonstrates that analyzing correlations in datasets of stochastically generated wavefunction snapshots provides useful probes of ergodicity-breaking phenomena. Our protocol can be readily applied to characterize quantum dynamics in ongoing experiments~\cite{lunkin2026}. The advantage of our protocol is that it requires only a moderate number of snapshots, i.e., $N_s\ll\mathcal{D}$, while the measurements relying on predefined observables, such as the imbalance~\cite{bloch_dalibard_12,schreiber_hodgman_15,choi_hild_16,Bordia17,li2025,hur2025}, transition probabilities~\cite{Yao23,lunkin2026}, or survival probabilities~\cite{lunkin2026}, require much larger amount of snapshots~\cite{lunkin2026}.

\acknowledgements 
R.~A. and M.~D. thank Santiago Acevedo, Markus Heyl, Alessandro Laio, and the Google quantum AI team for inspiring discussions. 
R.~A. and D.~S.~B. also thank Roberto Verdel for useful discussions and collaborations on related projects.
M.~D. was partly supported by the QUANTERA DYNAMITE PCI2022-132919, by the EU-Flagship programme Pasquans2, by the PNRR MUR project PE0000023-NQSTI, the PRIN programme (project CoQuS), and by the ERC Consolidator grant WaveNets (Grant agreement ID: 101087692). 
The work was also partially supported by INFN through the project QUANTUM (M. D.).
L.V. acknowledges support from the Slovenian Research and Innovation Agency (ARIS), Research core funding Grants No.~P1-0044, N1-0273 and J1-50005, as well as the Consolidator Grant Boundary-101126364 of the European Research Council (ERC). M. H. acknowledges support from the Polish National Agency for Academic Exchange (NAWA)’s Ulam Programme (project BNI/ULM/2024/1/00124). We gratefully acknowledge the High Performance Computing Research Infrastructure Eastern Region (HCP RIVR) consortium~\cite{vega1} and European High Performance Computing Joint Undertaking (EuroHPC JU)~\cite{vega2}  for funding this research by providing computing resources of the HPC system Vega at the Institute of Information sciences~\cite{vega3}. The computation of the BID is performed using the open source package dadapy~\cite{dadapy}.

% --- Appendices ---

\appendix

\section{Details on numerical implementation of the quantum sun model }\label{app:QSM} 

\subsection{Quantum sun model without $U(1)$ symmetry}

We first review the quantum sun model without $U(1)$ symmetry, as formulated in Refs.~\cite{suntajs_vidmar_2022,suntajs_deroeck_24}. It consists of a collection of $L$ qubits, or spins $1/2$, divided into two subsystems. The first subsystem, dubbed the ergodic dot, contains $N$ qubits that interact with each other through all-to-all couplings. The second subsystem consists of $L'=L-N$ additional qubits that do not mutually interact directly. The Hamiltonian is:
\begin{equation}\label{eq:qsun}
H = H_{\mathrm{dot}} + g_{0}\sum_{j=0}^{L'-1} \alpha^{u_j} 
S^x_{n_{(j)}} S^x_j
+ \sum_{j=0}^{L'-1} h_j S^z_j.
\end{equation}
Each outer qubit $j$ is coupled only to one randomly selected qubit $n_{(j)}$ from the ergodic dot. The former is located at a distance $u_j$ from the dot, where $u_j$ is drawn uniformly from the interval $[j - \zeta_j,\, j + \zeta_j]$ with $\zeta_j$ for $j>0$ and $u_j=0$ for $j=0$. Its coupling strength to the corresponding dot's qubit is given by $g_0 \, \alpha^{u_j}$. In addition, each outer qubit experiences a random magnetic field $h_j$ drawn uniformly from the range $[1 - W,\, 1 + W]$. 

The internal degrees of freedom of the sun, i.e., the dot, are modeled by a random matrix. In the standard binary ordering of the basis~\cite{suntajs_deroeck_24}, the matrix of the dot Hamiltonian can be written as
\begin{equation}
H_{\mathrm{dot}} =\gamma \frac{R'_\mathrm{dot}}{\sqrt{2^{N}+1}} \otimes \mathbb{I} 
\end{equation}
where $R'_\mathrm{dot}= (A + A^{T})/\sqrt{2}$ and $A$ is a real matrix whose elements $A_{ij}$ are independently sampled from a normal distribution, i.e., $R'_\mathrm{dot}$ corresponds to a matrix from the Gaussian Orthogonal Ensemble (GOE) of matrices of dimension $2^N \times 2^N$. 
The normalization follows Ref.~\cite{suntajs_deroeck_24}, where the Hilbert-Schmidt norm of the dot Hamiltonian was set to $||H_{\mathrm{dot}}||=\gamma$, with $||O||^{2}={\rm Tr}(O^2)/\mathcal{D}^{\mathrm{all}}$. Here,
$\mathcal{D}^{\mathrm{all}}$ is the sum of the dimensions of all
magnetization sectors, which coincides with the dimension of the full space of configurations of $L$ qubits, i.e., $\mathcal{D}^{\mathrm{all}}=\mathcal{D}_0=2^L$. The normalization is ensured, for large $N$, by the normalization factor $\sqrt{2^{N}+1}$. Alternatively, numerical normalization can be adopted, i.e., $H_{\mathrm{dot}} =\gamma \frac{R'_\mathrm{dot}\otimes\mathbb{I}}{||R'_\mathrm{dot}\otimes\mathbb{I}||}$.

The parameter $\alpha$ tunes the system across the ergodicity-breaking transition. Based on theoretical arguments of thermal avalanches~\cite{deroeck_huveneers_17,luitz_huveneers_17}, the ergodic dot thermalizes the entire system for $\alpha > \alpha_c^{\rm theory} = 1/\sqrt{2}$ while the avalanche stalls for $\alpha < \alpha_c^{\rm theory}$ and ergodicity is broken.  The transition is clearly visible in finite-size numerical calculations~\cite{suntajs_vidmar_2022,suntajs_deroeck_24}, with the transition ${\alpha}_c$ being quantitatively close to the theoretical prediction $\alpha_c^{\rm theory}$.

\subsection{Quantum sun model with $U(1)$ symmetry}

In our study, we consider the $U(1)$ symmetric variant of the quantum sun model, introduced in Ref.~\cite{pawlik_zakrzewski_2024}. We modify it with the canonical normalization condition of Ref.~\cite{suntajs_deroeck_24} for the dot Hamiltonian.

The interaction between the dot qubits and the outer qubits becomes $U(1)$ symmetric by adding $S^yS^y$ terms to the interaction term in Eq.~\eqref{eq:qsun}. Then, the $U(1)$ symmetric Hamiltonian is given by Eq.~\eqref{eq:qsun:U1} where, in addition, the Hamiltonian $H_{\mathrm{dot}}^{U(1)}$ is different from $H_{\mathrm{dot}}^{}$ as it also satisfies the $U(1)$ symmetry. We adopt the same normalization condition for $H_{\mathrm{dot}}^{U(1)}$ as for $H_{\mathrm{dot}}$ in Ref.~\cite{suntajs_deroeck_24}, i.e., $||H_{\mathrm{dot}}^{U(1)}||=\gamma$, where, however, the Hilbert-Schmidt norm $||O||^{2}={\rm Tr}(O^2)/\mathcal{D}$ is now related to the Hilbert space of dimension $\mathcal{D}$ corresponding to a chosen $U(1)$ sector. To give an example, for the zero magnetization sector considered in our study, $\mathcal{D}=\binom{L}{L/2}\propto 2^{L}/\sqrt{L}$, and thus the operator $H_{\mathrm{dot}}^{U(1)}$ is represented by a matrix of size $\mathcal{D}\times\mathcal{D}$.

Practically, we first generate a matrix ${R}_{\mathrm{dot}}^{U(1),\mathrm{all}}$, which is a unification for all $U(1)$ sectors. To do so, we start with the matrix $R'_\mathrm{dot}$ where we set to zero the elements of this matrix that correspond to the processes between the basis states of the dot which do not conserve the $U(1)$ symmetry, i.e., $R'_\mathrm{dot} \rightarrow {R'}^{U(1)}_\mathrm{dot}$. Next, assuming the standard binary ordering of the basis~\cite{suntajs_deroeck_24}, we embed ${R'}^{U(1)}_\mathrm{dot}$ into the full matrix, i.e., ${R}_{\mathrm{dot}}^{U(1),\mathrm{all}} = {R'}^{U(1)}_\mathrm{dot}\otimes \mathbb{I}$. The operator ${R}_{\mathrm{dot}}^{U(1),\mathrm{all}}$ is represented by a matrix of size $\mathcal{D}^{\mathrm{all}}\times\mathcal{D}^{\mathrm{all}}$ where $\mathcal{D}^{\mathrm{all}} =2^L$.

As the system is $U(1)$ symmetric, the matrix ${R}_{\mathrm{dot}}^{U(1),\mathrm{all}}$ is block-diagonal with respect to the magnetization sectors. Therefore, after selecting a magnetization sector, we reduce the operator ${R}_{\mathrm{dot}}^{U(1),\mathrm{all}}$ to the corresponding Hilbert space of that sector, i.e.,
${R}_{\mathrm{dot}}^{U(1),\mathrm{all}}   \rightarrow {R}_{\mathrm{dot}}^{U(1)}$. Here, $R_{\mathrm{dot}}^{U(1)}$ is an operator in the reduced Hilbert space of dimension $\mathcal{D}$. Finally, we construct the dot operator as
\begin{equation}
\label{eq:qsun:U1norm}
H_{\mathrm{dot}}^{U(1)}  =\gamma \frac{{R}_{\mathrm{dot}}^{U(1)}}{\Big|\Big|{R}_{\mathrm{dot}}^{U(1)}\Big|\Big|},
\end{equation}
where the normalization is implemented numerically.

As the dimension of the zero magnetization sector studied here scales exponentially with $L$, i.e., $\mathcal{D}\propto 2^{L}/\sqrt{L}$, the system with $U(1)$ symmetry is expected to exhibit behavior similar to that of the system without $U(1)$ symmetry. Namely, for $\alpha > \alpha_c^{\rm theory}$, the dot thermalizes the entire system, whereas for $\alpha < \alpha_c^{\rm theory}$, ergodicity is broken. The transition in the $U(1)$ symmetric quantum sun model for the zero magnetization sector was studied numerically in Ref.~\cite{pawlik_zakrzewski_2024} with the estimated value $\alpha_c\approx0.76$. Since we implement the model from Ref.~\cite{pawlik_zakrzewski_2024} modified with the canonical normalization condition of Ref.~\cite{suntajs_deroeck_24}, we also provide an estimate of the transition point in the modified model in the next section.

\section{Ergodicity braking transition point in $U(1)$ symmetric quantum sun model} 
\label{app2}

\subsection{Transition point from the spectral measures}

We estimate the value of the ergodicity breaking transition point for the Hamiltonian in Eq.$~$\eqref{eq:qsun:U1} with the dot normalization given by Eq.$~$\eqref{eq:qsun:U1norm} for a parameter set: $N=3$, $\gamma=1$, $g_0=1$, $\zeta=0.2$, $W=0.5$. To this end, we employ the spectral measures~\cite{suntajs_vidmar_2022,suntajs_deroeck_24}, i.e., the mean gap ratio and the spectral form factor.

We first analyze the mean gap ratio $r$~\cite{oganesyan_huse_07}. The ratio $\tilde{r}_\mu$ is defined as
\begin{equation}\label{eq:defr}
    r_\mu = \frac{\min\{\delta E_\mu, \delta E_{\mu-1}\}}{\max\{\delta E_\mu, \delta E_{\mu-1}\}} \;,
\end{equation}
where $\delta E_\mu = E_{\mu+1} - E_\mu$ is the level spacing between the eigenlevels $\mu$ and $\mu-1$, respectively. We target eigenlevels $\mu$ close to the mean energy $\bar E={\rm Tr}(H)/\mathcal{D}$. By definition, $\{\tilde{r}_\mu\}$ values fall in the interval $[0, 1]$, and hence no unfolding procedure is needed to eliminate the influence of finite-size effects through the local density of states. To obtain the average value $r$,
\begin{equation} \label{eq:def_r}
    r = \langle \langle r_\mu \rangle_\mu \rangle_H \;,
\end{equation}
we first average over $N_{\rm eig}=500$ eigenstates near the center of the spectrum for each Hamiltonian realization, denoted by $\langle \cdots \rangle_\mu$ in Eq.~(\ref{eq:def_r}), and then over an ensemble of spectra for different Hamiltonian realizations, denoted by $\langle \cdots \rangle_H$. In the ergodic phase, $r$ assumes the GOE value $r_{\rm GOE}\approx 0.5307$~\cite{atas_bogomolny_13}, while the prediction for energy levels with Poissonian statistics is $r_{\rm Poisson} = 2\ln 2 - 1 \approx 0.3863$~\cite{oganesyan_huse_07}. 

\begin{figure}
    \centering
    \includegraphics[width=\linewidth]{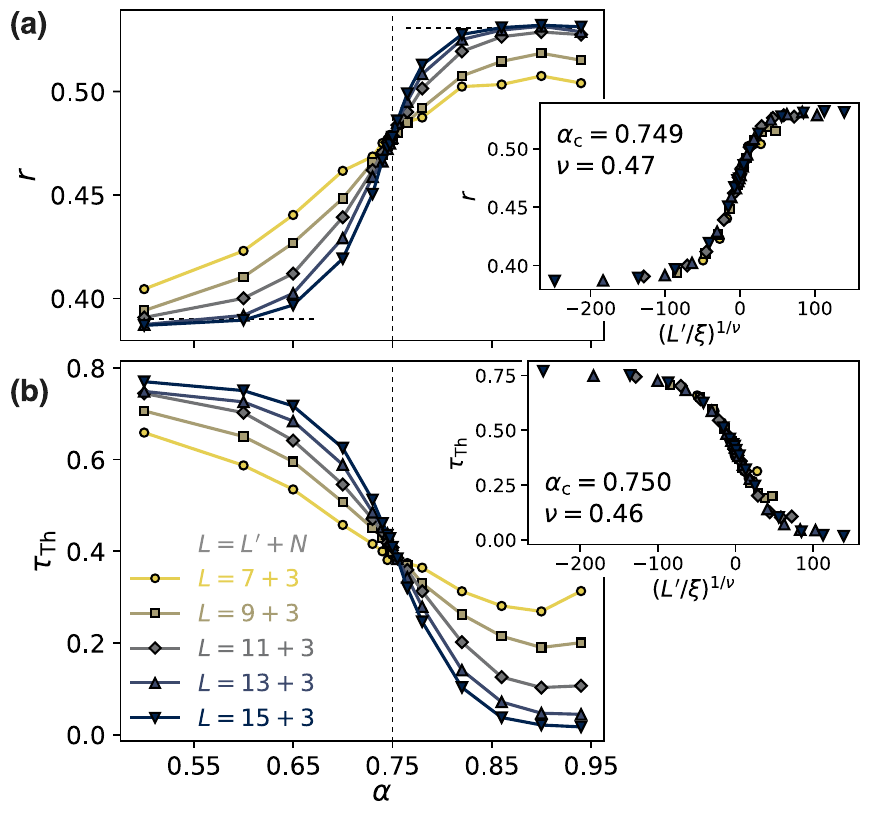}
    \caption{
    \textbf{Transition point estimation from spectral measurements.} 
    Gap ratio $r$ {\bf(a)} and the Thouless time normalized by the Heisenberg time,
    $\tau_{\rm Th}=t_{\rm Th}/t_{\rm H}$, {\bf (b)}, as the spectral indicators of the ergodicity-breaking transition. The transition point is estimated from scaling collapses in the insets as $\alpha_c\approx 0.75$, see the black vertical dashed lines in the main panels. Horizontal dashed lines in {\bf(a)} represent the theoretical limits of $r$ for the ergodic phase ($r=0.5307)$ and for the non-ergodic phase ($r=0.3863)$.}
    \label{fig:qs_gap_ratio}
\end{figure}

In Fig.$~$\ref{fig:qs_gap_ratio}(a), we show the average gap ratio $r$ as a function of the tuning parameter $\alpha$. We observe a flow to the Poisson and GOE statistics for $\alpha<0.74$ and $\alpha >0.76$, respectively, with a crossing point developing around $\alpha=0.75$. 
To estimate $\alpha_c$, we perform the finite size scaling collapse of the data from Fig.$~$\ref{fig:qs_gap_ratio}(a) with the ansatz of Ref.~\cite{swietek_hopjan_25}, i.e., we consider the correlation length $\xi=[\ln(\alpha/\alpha_c)^2]^{-\nu}$, where $\nu$ is the critical exponent, and plot the results as a function of $(L'/\xi)^{1/\nu}$. We consider $L'=L-N$ instead of $L$ to remove the finite size effects of the dot. The optimal scaling collapse, which is given by the minimization of the cost function~\cite{suntajs_bonca_20b}, is obtained for $\alpha_c^{r} \approx 0.749$ and $\nu^{r} \approx 0.47$, see the inset of Fig.$~$\ref{fig:qs_gap_ratio}(a). We note that while $\nu=1$ is expected in the thermodynamic limit, as argued in Ref.~\cite{swietek_hopjan_25}, the results in finite systems may typically give rise to $\nu<1$~\cite{swietek_hopjan_25}, which is also observed here.

We further consider spectral properties beyond the short-range spectral correlations and study the long-range spectral correlation via the spectral form factor (SFF)$~$\cite{suntajs_bonca_20a, Sierant2020, suntajs_prosen_21, suntajs_vidmar_2022,hopjan_vidmar_23b,Jiricek_26}. We follow the standard unfolding procedure, i.e., we introduce the cumulative spectral function $\mathcal{G}(E) = \sum_{\mu}\Theta(E-E_{\mu, H})$, where $\Theta$ is the unit step function. The stepwise distribution function is then smoothed out by fitting a polynomials $g_m(E)$ of degree $m$ to $\mathcal{G}(E)$ and the unfolded eigenvalues are defined as $\tilde{E}_{\mu, H} = g_m(E_{\mu, H})$. We set $m=3$ in our study. With the unfolded spectrum $\{\tilde{E}_{\mu, H} \}$ at hand, we construct the normalized SFF$~$\cite{suntajs_bonca_20a},
\begin{equation}
\tilde{K}^{}_n(\tau) = \frac{1}{Z}\Bigg\langle \Bigg| \sum_{\mu=1}^{\cal D} \rho( \tilde{E}_{\mu, H}) e^{-i 2\pi\tilde{E}_{\mu, H} \tau} \Bigg|^2 \Bigg\rangle_H,
\label{eqn:K_uf}
\end{equation}
where $ \tau$ is the scaled time, $Z=\langle\sum_{\mu}|\rho(\tilde{E}_{\mu, H})|^2\rangle_H$ is the normalization such that $\tilde{K}^{}_n( \tau \gg1)\approx 1$ and $\rho(\tilde{E})$ is a filter function that eliminates the contributions of spectral edges. We note that each of the unfolded spectra is filtered separately. We use the Gaussian filter that was used in recent studies$~$\cite{suntajs_bonca_20a, Sierant2020, suntajs_prosen_21, suntajs_vidmar_2022,hopjan_vidmar_23b,Jiricek_26}, defined as $\rho(\tilde{E}_{\mu, H}) = \exp\{-\frac{(\tilde{E}_{\mu, H}-\bar{E}_{H})^2}{2(\eta \Gamma_{H})^2}\}$, where $\bar{E}_{H}$ and $\Gamma_{H}^2$ are the average energy and the energy variance, respectively, at a given Hamiltonian realization. A dimensionless parameter $\eta$ controls the effective fraction of eigenstates included in $\tilde{K}^{}_n(\tau)$.  We set $\eta=0.7$ in our study.

The onset of universal behavior, i.e., the agreement with the GOE prediction $\tilde{K}^{{\rm GOE}}_{n}(\tau) = 2\tau-\tau\ln{(2\tau+1)}$, emerges at times larger than the Thouless time $\tau_{{\rm Th}}$$~$\cite{suntajs_bonca_20a, Sierant2020,suntajs_prosen_21, suntajs_vidmar_2022, hopjan_vidmar_23b, Jiricek_26}. To extract the value of $\tau_{\rm Th}$ (note that it is expressed in dimensionless units due to spectral unfolding), we calculate $\tilde{K}^{}_n(\tau)$ for $N_{\tau}=5000$ discrete time steps $\tau_i$ in the logarithmic grid and smooth out fluctuations in $\tilde{K}^{}_n(\tau)$ by calculating a running mean over $300$ values of $K_n(\tau_i)$. Finally, we interpolate the smoothen normalized SFF with $10000$ data points. With the interpolated results at hand, we compute the difference between $\tilde{K}^{}_n(\tau)$ and the GOE prediction, using a deviation measure defined as~\cite{suntajs_bonca_20a}
\begin{equation}
    \Delta \tilde{K}(\tau) = \Bigg|\ln\Bigg(\frac{\tilde{K}^{}_n(\tau)}{\tilde{K}^{{\rm GOE}}_{n}(\tau)}\Bigg)\Bigg|;\hspace{3mm} \Delta \tilde{K}(\tau_{{\rm Th}}) = \epsilon. 
    \label{eqn:K_dev} 
\end{equation}
Here, the scaled Thouless time $\tau_{{\rm Th}}$ is the time at which $\Delta K(\tau)$ becomes greater than a chosen threshold value $\epsilon$. We note that although the absolute value of the extracted Thouless time depends on the threshold $\epsilon$, a power-law exponent characterizing the system size dependence of the Thouless time is expected to be independent of this parameter. We have checked (not shown) that the exponent of the Thouless time dependence does not change significantly for $\epsilon=0.1-0.9$. We choose $\epsilon=0.2$ in further analysis.

In Fig.$~$\ref{fig:qs_gap_ratio}(b), we show the normalized Thouless time $\tau_{{\rm Th}}$ as a function of the tuning parameter $\alpha$. We observe that $\tau_{{\rm Th}}\rightarrow 0$ for $\alpha >0.76$ while for $\alpha <0.74$ the normalized  Thouless time $\tau_{{\rm Th}}$ grows with the system size $L$. The transition point is signaled by a system-size independent normalized Thouless time $\tau_{{\rm Th}}$ which emerges close to $\alpha\approx0.75$. In physical units, this means that the Thouless time $t_{\rm Th}$ exhibits the same scaling with the system size $L$ as the Heisenberg time, i.e., $t_{\rm Th}(L)\propto t_{\rm H}(L)$. 

As in the case of the gap ratio, we perform the finite-size scaling collapse of the results with the ansatz of Ref.~\cite{swietek_hopjan_25}. The optimal scaling collapse, given by the minimization of the cost function~\cite{suntajs_bonca_20b}, is obtained for $\alpha_c^{\rm SFF} \approx 0.750$ and $\nu^{\rm SFF} \approx 0.46$; see the inset of Fig.$~$\ref{fig:qs_gap_ratio}(b). These values are close to those estimated from the gap ratio statistics. We therefore conclude that both indicators gives consistently $\alpha_c^{}\approx \alpha_c^{r}\approx \alpha_c^{\rm SFF}\approx 0.75$.

\subsection{Transition point from participation entropies}

\begin{figure}
    \centering
    \includegraphics[width=\linewidth]{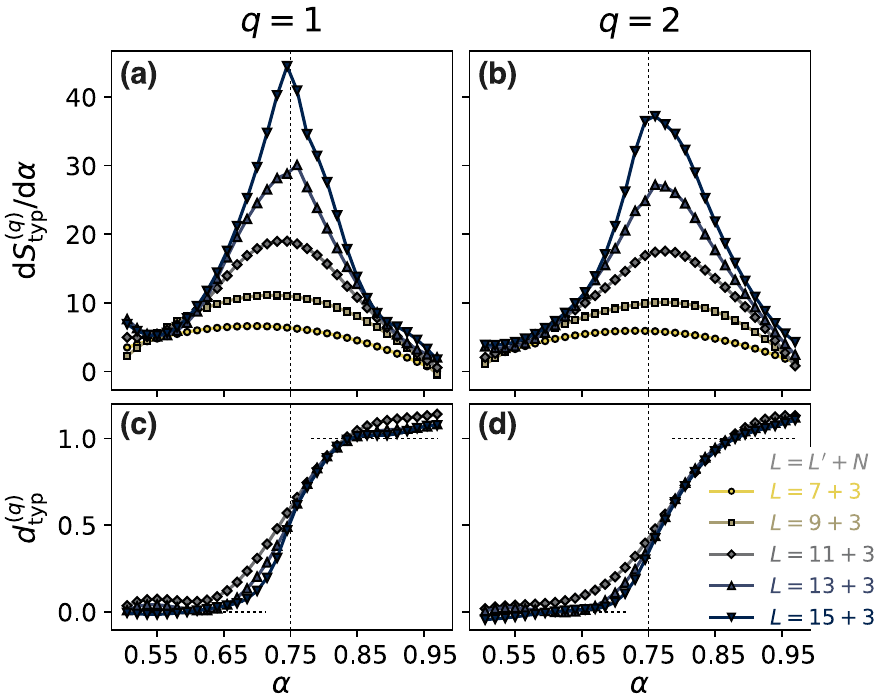}
    \caption{ \textbf{Participation entropy and floating fractal dimension.} 
    {\bf (a) and (b)}
    Derivative of the typical participation entropy $S_{\rm typ}^{(q)}$ with respect to $\alpha$ as the wavefunction indicators of the ergodicity breaking transition, shown for $q=1$ and $q=2$, respectively.
    {\bf (c) and (d)} 
    The corresponding results for the floating fractal dimension $d^{(q,L)}_{\rm typ}$ for $q=1$ and $q=2$, respectively.
    Black vertical dashed lines indicate the predicted transition point $\alpha_c \approx 0.75$. 
    }
    \label{fig:qs_l_dependence}
\end{figure}

We here show that the ergodicity breaking transition is also reflected in the properties of mid-spectrum eigenstates. 
To this end, we study the scaling of the participation entropies~\cite{suntajs_deroeck_24}, which characterize the amount of delocalization in a chosen basis. 

We consider the inverse participation ratio (IPR), generalized to an arbitrary index $q$, for a selected number of Hamiltonian eigenstates $\ket{\mu}$, which is defined as
\begin{equation}\label{eq:def_IPR}
    P_q^{-1}(\ket{\mu}) = \sum_{I=1}^{\mathcal{D}}|\braket{I|\mu}|^{2q}\;.
\end{equation}
The states $\ket{I}$ are the product states in the computational basis in the zero magnetization sector of the $U(1)$ symmetry. Then, the averaged participation entropy is defined~\cite{suntajs_deroeck_24} as
\begin{equation}\label{eq:def_participation_entro}
    S_{\rm typ}^{(q)} = \frac{1}{1-q}\langle \langle \ln P_q^{-1} \rangle_\mu\rangle_{H}\;.
\end{equation}
For the value $q \to 1$, the participation entropy reduces to the von Neumann participation entropy,
\begin{equation} \label{eq:def_S1typ}
S_{\rm typ}^{(1)} = - \Big\langle\Big\langle \sum_{I=1}^{\mathcal{D}}|\braket{I|\mu}|^{2} \ln  |\braket{I|\mu}|^{2} \Big\rangle_\mu \Big\rangle_H\;.
\end{equation}
We refer to these participation entropies as the typical participation entropies, as the averaging of the IPRs are performed outside the logarithm. 

The participation entropy was shown~\cite{suntajs_deroeck_24}, for the quantum sun model without $U(1)$ non-symmetry, to be an increasing function of the system size in the ergodic phase and at criticality, and to be saturating in the non-ergodic phase due to Hilbert-space localization of the wavefunction. This suggests that the slope of the participation entropy as a function of $\alpha$ is maximal at the transition point, which can be quantified by the derivative of the participation entropy with respect to $\alpha$, i.e., $dS_{\rm typ}^{(q)}/d\alpha$~\cite{suntajs_deroeck_24}.

In Figs.$~$\ref{fig:qs_l_dependence}(a) and~\ref{fig:qs_l_dependence}(b), we show examples of such derivatives for the $U(1)$ conserving quantum sun model at $q=1$ and $q=2$, respectively. The derivatives $dS_{\rm typ}^{(q)}/d\alpha$ develop divergent peaks at the predicted transition point $\alpha_c \approx 0.75$. As a technical remark, we note that the derivatives are calculated by first interpolating the raw data for $S_{\rm typ}^{(q)}$ with cubic splines, and then evaluating the first derivatives of the interpolated data~\cite{suntajs_deroeck_24}.

We further consider the scaling of the participation entropy with the system size $L$. To this end, we assume a scaling ansatz~\cite{suntajs_deroeck_24} 
\begin{equation} \label{eq:def_Sq_dq}
S_{\rm typ}^{(q)} = d_{\rm typ}^{(q)} \ln{ \mathcal{D}} + b_{\rm typ}^{(q)}\;,
\end{equation}
where $d_{\rm typ}^{(q)}$ is interpreted as the wavefunction fractal dimension. In a finite system outside of the asymptotic regime, the parameters $d_{\rm typ}^{(q)}$ and $b_{\rm typ}$ are additionally system size dependent. To account for the finite-size effects, we define a floating fractal dimension~\cite{suntajs_deroeck_24},
\begin{equation} \label{eq:def_dqL}
d^{(q,L)}_{\rm typ}= \frac{S_{\rm typ}^{(q)}(L)-S_{\rm typ}^{(q)}(L-1)}{\ln[{D}(L)]-\ln[{D}(L-1)]}\;.
\end{equation}
In Figs.$~$\ref{fig:qs_l_dependence}(c) and~\ref{fig:qs_l_dependence}(d), we show examples of the floating fractal dimension for $q=1$ and $q=2$, respectively. We observe that for $\alpha<\alpha_c$, the fractal dimension flows towards zero. For $\alpha>0.85$, we observe $d^{(q,L)}_{\rm typ}>1$, where $d^{(q,L)}_{\rm typ}\rightarrow 1$ with increasing system size $L$, consistent with the full delocalization of the wavefunction. For $\alpha_c<\alpha<0.85$, the system size dependence of the floating fractal dimension, as well as its fate in the thermodynamic limit, is less clear. That being said, we expect the wavefunction to be multifractal at criticality. The results in Figs.$~$\ref{fig:qs_l_dependence}(c) and~\ref{fig:qs_l_dependence}(d) confirm this expectation, as we observe rather stable values $0<d^{(q,L)}_{\rm typ}<1$ at the predicted transition point $\alpha_c\approx0.75$.

\bibliographystyle{biblev1}
\bibliography{ref, references, references1, references2, references3}

\begin{thebibliography}{10}
\bibitem{PCA_footnote}
\bibinfo{title}{{PCA} is a dimensionality reduction method that projects a dataset onto a new set of orthogonal axes, the principal components, which are ordered by the amount of variance. here, it is used to visualize high-dimensional wave-function snapshot datasets in a $3$-dimensional space aligned with the dataset's $3$ orthogonal directions that contain the most variance.}
\end{thebibliography}

\end{document}